\documentclass[12pt,notoc]{JHEP}

\newcommand{\be}{\begin{equation}}
\newcommand{\ee}{\end{equation}}
\newcommand{\bea}{\begin{eqnarray}}
\newcommand{\eea}{\end{eqnarray}}
\newcommand{\bref}[1]{(\ref{#1})}
\newcommand{\br}{\mathrm{b}}
\newcommand{\Lagb}{\mathcal{L}_\br}
\newcommand{\Lag}{\mathcal{L}}
\newcommand{\rhob}{\rho_\br}
\newcommand{\pb}{p_\br}
\newcommand{\const}{\mathrm{const.}}
\newcommand{\av}[1]{\left<{#1}\right>}
\newcommand{\C}{\mathcal{C}}

\title{\bf Brane Cosmology Solutions with Bulk Scalar Fields}

\author{Stephen C. Davis\\ 
LPT, Universit\'e Paris--XI, 
B\^atiment 210, F--91405 Orsay Cedex, France \\ 
{E-mail:} \email{Stephen.Davis@th.u-psud.fr} }

\abstract{
Brane cosmologies with static, five-dimensional and $Z_2$ symmetric
bulks are analysed. A general solution generating mechanism is
outlined. The qualatitive cosmological behaviour of all such solutions
is determined. Conditions for avoiding naked bulk singularities are
also discussed. The restrictions placed on the solutions by the
assumption of such a static bulk are investigated. In particular the
requirement of a non-standard energy-momentum conservation law. The
failure of such solutions to provide viable quintessence terms in the
Friedmann equations is also discussed.
}


\preprint{LPT--Orsay/01--113}

\begin{document}

\section{Introduction}

The possibility that our four-dimensional universe may be embedded in a
higher dimensional `bulk' spacetime has received a great deal of
attention from cosmologists. This idea is motivated by the existence
of D-branes in string/M-theory. Our universe could then be one of these
branes, moving through a non-compact higher dimensional space. In
general, effects arising from the extra dimensions in this setup will be
incompatible with the observed behaviour of our universe. However, if 
a negative cosmological constant is added to the bulk, it is possible (to
leading order) to re-obtain standard four-dimensional gravity~\cite{RS} and
cosmology on the brane~\cite{BWcos}. 

Early work concentrated on models with constant bulks, however in the
context of string/M-theory it is natural for bulk scalar fields to be
present too, and so their effects should not be
neglected. It is also possible that the variation of these fields could
provide new mechanisms for previously unexplained cosmological
phenomena. For example, they may produce effective quintessence terms in
the brane Friedmann equation. This could then account for the observed
acceleration of our universe's expansion.

By projecting the higher dimensional curvature tensors onto the brane,
and applying the relevant boundary conditions it is possible to obtain
four-dimensional Einstein equations~\cite{proj}. In the case of brane
cosmologies with constant bulk energy densities, this approach allows
the evolution of the fields on the brane to be completely determined. 
When scalar fields are introduced, the behaviour of
the projected Weyl tensor is no longer fully determined by the boundary
conditions. We therefore need more information about the bulk before we
can find the effective Einstein equations on the brane.

One possibility is to to make assumptions about the bulk solution near
the brane~\cite{dLconf,dLp}, although we will not know if they are
justified or not. A preferable option is to solve the full higher
dimensional Einstein equations. So far no general solutions have been
found. However solutions can be found in a few special cases in which
the equations simplify. For example, we could assume the bulk energy
density has the form of a perfect fluid~\cite{fluidbulk}. Alternatively,
if the bulk potential energy resembles that of a supergravity theory,
BPS solutions can be found~\cite{SGanne}.

In this paper I will consider models with just one scalar field and a
single 3-brane in a five-dimensional bulk. I will be mainly concerned
with $Z_2$ symmetric spacetimes (i.e.\ the solution on one side of the
brane is a reflection of that on the other), although a few results
apply to the more general case. To simplify the analysis I will only
consider solutions with static bulks. The position of the brane will still
be time-dependent, and it is from the brane's movement through the bulk
that the cosmological evolution of our universe arises.  This
generalises an idea by Ida~\cite{Ida}. It turns out that the BPS
solutions considered elsewhere are a subset of those considered in this
paper.

In section~\ref{s:back} of this paper I will outline the model used,
and relate the brane Friedmann equation to the bulk solutions. The
assumption of having a static bulk places restrictions on the
brane's evolution, which I will discuss. In section~\ref{s:sol} I will
show how the static assumption allows the field equations to be simplified,
and how they provide a way of generating solutions. The qualitative
cosmological properties of all such solutions are discussed in
sections~\ref{s:gen} and \ref{s:sg}. Section~\ref{s:sg} deals with a
special case of conformally flat solutions, which have been more widely
discussed in the literature~\cite{SGsol}. For illustration, a particular set
of solutions is presented in section~\ref{s:eg}. Finally the results are
summarised in section~\ref{s:sum}.

\section{Static Bulk Brane World Cosmology}
\label{s:back}

For a 3-brane in a 5-dimensional spacetime, the 4-dimensional effective
Einstein tensor $\widehat G_{\mu\nu}$ is related to a projection of the bulk
Einstein tensor~\cite{proj}.
\bea
&&\widehat G_{\mu\nu} = \frac{2}{3}\left(
G_{\lambda \sigma} \hat g^\lambda_\mu  \hat g^\sigma_\nu 
+\left(G_{\lambda \sigma} n^\lambda n^\sigma 
- \frac{1}{4}G^\sigma{}_\sigma \right)\hat g_{\mu\nu}\right) 
\nonumber \\ && \hspace{.5in}
- \frac{1}{2}\left(K_{\mu\lambda} - K \hat g_{\mu\lambda}\right)
\left(K^\lambda_\nu - K \hat g^\lambda_\nu\right) - E_{\mu\nu} \ ,
\eea
where 
\be
K_{\mu\nu} = \hat g^\lambda_\mu \hat g^\sigma_\nu 
\nabla_{\! (\lambda} n_{\sigma)}
\ee
is the extrinsic curvature of the brane,
\be
\hat g_{\mu\nu} = g_{\mu\nu} - n_\mu n_\nu
\ee
is the induced metric,
\be
E_{\mu\nu} = C^\alpha{}_{\lambda \beta \sigma} n_\alpha n^\beta 
\hat g^\lambda_\mu  \hat g^\sigma_\nu
\label{Edef}
\ee
is the electric part of the Weyl tensor, and $n^\mu$ is the brane's normal.

Both $G_{\mu\nu}$ and $K_{\mu\nu}$ are directly determined by the
theory's energy momentum tensor. When scalar fields are present,
$E_{\mu\nu}$ is not. Thus a full bulk solution is required.

Rather than look for a completely general solution, I will consider the
simpler problem of finding solutions with static bulks. The cosmological
evolution of the system then arises from the movement of the brane
through the bulk. I will also assume that the solution on the brane is
homogeneous and isotropic, as in the standard cosmology. In this case
the bulk metric can be written as 
\be
ds^2 = -f^2(r) h(r)dT^2 + r^2 \Omega_{ij} dx^i dx^j + \frac{dr^2}{h(r)}
\label{g5}
\ee
where $\Omega_{ij}$ is the three dimensional metric of space with
constant curvature $k=-1,0,1$. 

A suitable action for a brane world with one bulk scalar field is
\be
S=\int_\mathrm{bulk} d^5x \sqrt{\! -g}
\left\{\frac{1}{2\kappa^2}R - \frac{1}{2}(\nabla \phi)^2 - V(\phi) \right\}
+ \int_\mathrm{brane} d^4x {\textstyle \sqrt{\! -\hat g}} 
\left\{ \frac{2}{\kappa^2}\av{K} + \Lagb(\phi) \right\} \ .
\label{action}
\ee
The brane Lagrangian $\Lagb$ includes all the Standard model fields,
which are confined to the brane, and any brane contributions to the
potential of the bulk scalar field $\phi$. The
notation $[X] = X^+ - X^-$, $\av{X} = (X^+ + X^-)/2$, which denotes the
change and average of quantities across the brane, has been introduced.
$\kappa^2 = 8 \pi / M_5^3$, where $M_5$ is the fundamental
five-dimensional Planck mass.

The energy-momentum tensor of the bulk scalar field is
\be
T^\mu_\nu = \left( \nabla^\mu \phi \nabla_{\! \nu} \phi - \delta^\mu_\nu 
\{ {\textstyle \frac{1}{2}}(\nabla\phi)^2 + V(\phi)\} \right) \ .
\label{Tbulk}
\ee
Since we are considering a static bulk, $\phi = \phi(r)$.
Two components of the Einstein equation are then
\be
G^T_T = \frac{3}{r^2}\left( h + \frac{r h'}{2} -k\right)
= -\kappa^2\left(\frac{1}{2}h\phi'^2 + V\right)
\label{GTT}
\ee
\be
G^r_r = G^T_T + \frac{3 h f'}{rf}
= \kappa^2\left(\frac{1}{2}h\phi'^2 - V\right) \ ,
\label{Grr}
\ee
and the equation of motion of the scalar field is
\be
\left\{\frac{3 h}{r} + \frac{h f'}{f} + h'\right\} \phi' + h \phi''
=\frac{dV}{d\phi} \ .
\label{phi}
\ee
The remaining bulk Einstein equations need not be considered since the
Bianchi identities ensure that they are automatically satisfied.
Subtracting eq.~\bref{GTT} from eq.~\bref{Grr} we see that
\be
f = \exp\left(\frac{\kappa^2}{3}\int r\phi'^2 dr \right) \ .
\label{fsol}
\ee

I will now consider the boundary conditions of the bulk solution, which
arise from the presence of the brane. 
Its position is given by $r=a(t)$, $T = T_\br(t)$,
where $t$ is the cosmological time experienced on the brane. 
The induced metric is then
\be
ds_\br^2 = -dt^2 + a^2(t) \Omega_{ij} dx^i dx^j \ .
\label{ds4}
\ee
This is a generalisation of the solutions of Ida~\cite{Ida}.

The brane's tangent vector is $u^\mu = (\dot T,0,0,0,\dot r)$,
where dots denote differentiation by $t$.  This gives the
normalisation condition
\be
-f^2 h^2 \dot T^2 + \dot r^2 = -h \ .
\label{norm}
\ee
The unit 1-form normal to the brane is then
$n_\mu = (-\dot r,0,0,0, \dot T) f$. I will take $\dot T_\br$ and
$f$ to be positive, so that $t$ and $T$ have the same direction, and
$n_\mu$ is an outward normal.

The non-vanishing components of $K_{\mu\nu}$ are
\be
K_{\mu\nu}u^\mu u^\nu = -\frac{\ddot r + h'/2}{fh \dot T} 
- (\ln f)' fh \dot T
\ee
\be
K^i{}_j = \frac{fh \dot T}{r} \delta^i_j
\ee
For the metrics such as \bref{g5}, the spatial components of 
$E_{\mu\nu}$~\bref{Edef} satisfy
\be
E^i{}_j= \frac{1}{6} \left(\frac{1}{3}G^i{}_i 
- G^0{}_0 - G^5{}_5 \right) \delta^i_j + \frac{h-k}{r^2} \delta^i_j \ .
\label{Eij}
\ee
The only other non-zero component of the projected Weyl tensor is 
$E_{\mu\nu} u^\mu u^\nu = E^i{}_i$.

The action \bref{action} implies the following junction conditions
\be
2\av{n^\mu  \nabla_{\! \mu} \phi} = 
\frac{\delta \Lagb}{\delta \phi}
\label{jump1}
\ee
\be
2\av{K_{\mu\nu} - K \hat g_{\mu\nu}} = -\kappa^2 S_{\mu\nu}
\label{jump2}
\ee
on the brane, where $S_{\mu\nu}$ is the energy momentum tensor
corresponding to $\Lagb$.

For a perfect fluid 
$S^{\mu\nu} = (\rhob+\pb)u^\mu u^\nu + \pb \hat g^{\mu\nu}$.
The spatial part of the junction condition \bref{jump2} then implies
\be
\av{fh \dot T} = -\frac{\kappa^2}{6} a \rhob \ .
\ee
By considering the change in the normalisation condition \bref{norm} across the
brane we find
\be
\left[fh \dot T\right] = -\frac{3[h]}{\kappa^2 a \rhob} \ .
\ee
The average of eq.~\bref{norm} gives the Friedmann
equation on the brane
\be
\left(\frac{\dot a}{a}\right)^2 = -\frac{k}{a^2} + \av{U} 
+ \frac{\kappa^4}{36} \rhob^2 +  \frac{9[U]^2}{4\kappa^4 \rhob^2} \ ,
\label{Fried1}
\ee
where I have introduced
\be
U= -\frac{h-k}{r^2} \ .
\label{Udef}
\ee

I will now examine the conservation of energy-momentum on the brane.
The Codacci equation relates a projection of the bulk Einstein tensor to
the extrinsic curvature
\be
n^\mu G_{\mu\nu} \hat g^\nu_\lambda = 
D_{\! \mu} K^\mu_\lambda - D_{\! \lambda} K \ .
\label{codacci}
\ee
Combining this with $G_{\mu\nu} =\kappa^2 T_{\mu\nu}$ and the junction
condition \bref{jump2} gives the non-standard energy momentum
conservation equation on the brane
\be
\dot \rhob + 3 \frac{\dot a}{a}(\rhob + \pb) 
= - 2\dot a \av{fh \dot T \phi'^2 }  \ .
\label{emcon1}
\ee
The $\phi$ junction condition \bref{jump1} implies
\be
2\av{fh \dot T \phi'} = \frac{\delta \Lagb}{\delta \phi} \ .
\label{dLb}
\ee
If we assume a $Z_2$ symmetric bulk, the above two equations
simplify. With the help of the relation between $\phi'$ and $f$~\bref{fsol},
the energy conservation equation becomes
\be
\dot \rhob + 3 \frac{\dot a}{a}(\rhob + \pb) 
= \frac{\delta \Lagb}{\delta \phi} \dot \phi
= \left(-\frac{1}{f}\frac{d f}{d\phi}\rhob\right) \dot \phi \ .
\label{emcon2}
\ee

As with the usual brane world cosmology, the Friedmann equation
\bref{Fried1} has a $\rhob^2$ rather than a $\rhob$ term. To obtain
agreement with the standard cosmology at late times, $\rhob$ needs to be
composed of a brane tension part and an ordinary energy density
part. In contrast to the usual brane cosmology, we also need to satisfy
the non-standard energy conservation equation.
The simplest way to do this is to take 
\be
\rhob = (\lambda + \rho)/f 
\label{rhob}
\ee
and similarly $\pb = (-\lambda + p)/f$. The energy density
$\rho$ corresponds to the normal matter fields on the brane and obeys
the usual energy momentum conservation. The constant $\lambda$ is the
brane tension. The Friedmann equation then becomes
\be
\left(\frac{\dot a}{a}\right)^2 = -\frac{k}{a^2} + U + 
\frac{\kappa^4\lambda^2}{36f^2}
+ \frac{\kappa^4}{36 f^2} \rho^2 + \frac{\kappa^4 \lambda}{18f^2} \rho  \ .
\label{Fried2}
\ee
We see that the effective four-dimensional Einstein constant is
\be
8\pi G = \frac{\kappa^4 \lambda}{6 f^2(a)} \ ,
\ee
which is generally time dependent. If we require a vanishing effective
four-dimensional cosmological constant at late time, $U$ must tend to
$-\kappa^4/(36 f^2)$.

An alternative solution to
eq.~\bref{emcon2} is to take $\rhob = V_\br(\phi) + \rho$, where
$\rho$ obeys the standard conservation equation (or any other desired
law). The function $V_\br(\phi)$ must then be chosen to satisfy
eq.~\bref{emcon2}. Unfortunately this will require a brane potential
which changes form as the universe changes from radiation to matter
domination, so this type of solution is rather contrived.

In order for a solution of the form \bref{rhob} to be possible, the
brane matter fields must have $\phi$-dependent couplings in the
Lagrangian. There are various ways to do this. One possibility (i) is
to assume that brane matter is coupled to a metric which is conformally
related to $\hat g_{\mu\nu}$~\cite{dLconf}, so 
$\Lagb = \Lagb(\beta(\phi) g_{\mu \nu}, \ldots)$ where `$\ldots$'
represents all the matter fields which are confined to the
brane. Alternatively we could use a Lagrangian of the form
$\Lagb = \beta(\phi) \Lag_0(g_{\mu \nu}, \ldots)$,
$\Lag_0$ being $\phi$ independent. Exactly what the on-shell value of
$\Lagb$ is depends on the underlying theory. Two such possibilities are
(ii) $\Lagb = -2\pb$~\cite{dLp} and (iii) $\Lagb = 2\rhob$~\cite{dLrho}.
These three types of $\phi$-dependent Lagrangian have 
\be
\frac{\delta \Lagb}{\delta \phi}
= \frac{2}{\beta}\frac{d \beta}{d\phi} \left\{ \begin{array}{cc}
(\rhob -3 p_\br)/4 & \mathrm{(i)} \\  -p_\br & \mathrm{(ii)} \\
\rhob & \mathrm{(iii)} \end{array} 
\right.
\ee
For a cosmological constant ($p = -\rho$) these all imply 
$\beta \propto f^{-1/2}$. More generally when $p = w \rho$,  $\beta$
needs to be proportional to (i)
$f^{2/(3w-1)}$, (ii) $f^{-1/(2w)}$, or (iii) $f^{-1/2}$. Thus cases (i)
and (ii) require that energy densities with different equations of state
couple to $\phi$ differently. In addition, the above setup does not work
in case (i) for radiation ($w=1/3$) or in case (ii) for matter
($w=0$). Restrictions on the form of $\Lagb$ are also
discussed in refs.~\cite{david,restrict}.

\section{Solving the Field Equations}
\label{s:sol}

It is convenient to define
\be
\frac{\C}{r^4} = -\frac{1}{3}E^i{}_i
= \frac{r}{4f^2}\left(\frac{hf^2}{r^2}\right)' + \frac{k}{2r^2} 
\label{Cdef}
\ee
where the Einstein equations have been used to simplify the 
expression~\bref{Eij}.

Using the quantities $\C$ \bref{Cdef} and $U$ \bref{Udef}, the field
equations (\ref{GTT}--\ref{phi}) can be rearranged to give
\be
\C' = - \frac{\kappa^2}{3} \left(\C -\frac{k r^2}{2}\right) r\phi'^2
\label{Ceqk}
\ee
\be
U' + 4\frac{\C}{r^5} = 
- \frac{2\kappa^2}{3} \left(U - \frac{k}{r^2}\right) r\phi'^2
\label{Ueqk}
\ee
\be
\kappa^2 V = 6 U + \frac{3}{4} rU' - 3\frac{\C}{r^4} \ .
\label{Veqk}
\ee

If we take $\phi$ to be a constant we obtain the usual brane
cosmology solutions, with $f=1$, $\C$ constant, and
\be
U = -\frac{h-k}{r^2} = -\ell^2 + \frac{\C}{r^4} \ ,
\label{stdBW}
\ee
where $\ell$ is a constant. The Friedmann equation~\bref{Fried1} then agrees
with that derived elsewhere~\cite{Ida,Z2}.

For a given choice of $r\phi'$ or $\C$, eqs.~\bref{Ceqk} and \bref{Ueqk} can be
integrated. For example if we take $\kappa r\phi' = \alpha \sqrt{3}$,
with $\alpha$ constant,
\be
\C = \frac{k r^2}{2(2+\alpha^2)} + A_1 r^{-\alpha^2}
\label{Csol0}
\ee
\be
U = -\frac{\alpha^2(1+\alpha^2)}{(2+\alpha^2)(1-\alpha^2)} \frac{k}{r^2}
+ A_1\frac{4}{4-\alpha^2} r^{-4-\alpha^2} + A_2 r^{-2\alpha^2} \ .
\label{Usol0}
\ee
This extends the solution presented in refs.~\cite{david,reall}.

If $k=0$, the equations (\ref{Ceqk}--\ref{Veqk}) simplify to
\be
-\frac{\kappa^2}{3}r \phi' = \frac{1}{\C}\frac{d\C}{d\phi}
\label{Ceq}
\ee
\be
d \left(\frac{U}{\C^2}\right) = \frac{1}{\C} d (r^{-4}) 
\label{Ueq}
\ee
\be
V = -\frac{9}{2\kappa^4}
\frac{d\C}{d\phi}\frac{d}{d\phi}\left(\frac{U}{\C}\right)
+ \frac{6}{\kappa^2} U \ .
\label{Veq}
\ee
The expression for $f$~\bref{fsol} then reduces to $f \propto 1/\C$, and
so the Friedmann equation \bref{Fried2} becomes
\be
\left(\frac{\dot a}{a}\right)^2 =  U + \frac{\kappa^4}{36}\left(\lambda^2 + 
\rho^2 + 2\lambda\rho\right) \C^2  \ .
\label{Fried3}
\ee

Thus, given $\C(\phi)$, it is possible to determine the Friedmann
equation (from $U$), the potential $V$ as an explicit function of
$\phi$, and the evolution of $\phi$ with respect to $r$. Ideally we
would like to find all possible $U$ and $\C$ from an arbitrary
$V$. Unfortunately the non-linearity of the above equations makes this
practically impossible, although general solutions have been found for
exponential potentials~\cite{christos}.

The above results can also be derived using coordinates in which the
position of the brane is fixed. Following ref.~\cite{BWcos} we can take a
metric of the form
\be
ds^2 = -N^2(t,y) dt^2 + A^2(t,y) \Omega_{ij} dx^i dx^j + dy^2 \ .
\ee
The brane is fixed at $y=0$. If we then make the assumption that
$\phi = \phi(A)$, and define $h$ and $f$ as
\be
(\partial_y A)^2 - \left(\frac{\partial_t A}{N}\right)^2 = h(A)
\ee
\be
N \propto (\partial_t A) f(A) \ ,
\ee
the results of this paper will be re-obtained, with $r=A$. Using this
approach some of the results of section~\ref{s:sg} were found in
ref.~\cite{Me}. Other related work appears in ref.~\cite{pierreprep}.

\section{General Properties of Static Bulk Solutions}
\label{s:gen}

Using eq.~\bref{Ceq} I will now determine the qualitative features of
all $k=0$ brane world cosmologies with static bulks (except
for the special case $\C \equiv 0$, which is covered in the next
section). In general, $\phi$ simply rolls in the direction of decreasing
$|\C|$ as $r$ increases. More interesting things happen when
$d\C/d\phi=0$, $\C=0$ or $|\C| \to \infty$. The behaviour of the 
solutions near these points is easily seen using 
\be
\ln r = -\frac{\kappa^2}{3} \int \frac{\C d\phi}{(d\C/d\phi)} + \const
\ee
which is obtained from eq.~\bref{Ceq}.
If $d\C/d\phi \to 0$ then the integral diverges and so $\ln r \to \pm\infty$,
 depending on the sign of the integrand. Thus $r \to 0$ or
$\infty$. If $\C$ passes through zero then $dr/d\phi$ will change sign.
In other words the universe will stop expanding and start to
contract. If $\C$ diverges then $\ln r$ will either tend to a constant
or $-\infty$ depending on how $\C$ diverges. The other possibility 
($\ln r \to +\infty$) does not occur, since eq.~\bref{Ceqk} with $k=0$ implies
$d(\ln r)/d(\ln |\C|) \leq 0$, and so $r$ must decrease as $|\C| \to \infty$. 

It should be noted that the qualitative cosmological behaviour of the
above solution is not completely determined by eqs.~\bref{Ceq} and
\bref{Ueq}. We also need to consider the possibility of the $U$ term in
the Friedmann equation~\bref{Fried3} cancelling the other terms, to give
$\dot a = 0$.  In general this could occur at any value of $\C$ or
$d\C/d\phi$. Depending on the forms of $U$, $\C$ and $\rho$ it could
take an infinite amount of cosmological time to reach this point, or it
could be reached in finite time, after which the universe would start to
collapse. The subsequent evolution of the universe is then simply the
reverse of that up to the point where $\dot a =0$. This type of
evolution will prevent some values of $\phi$ being reached by certain
solutions.

We are mainly interested in cosmological solutions which
start at $r=0$ and end at $r=\infty$. The above analysis suggests we
want solutions which start at either a suitable turning point or
an infinity of $\C$. They should then have $\phi$ rolling down to another
turning point, without passing through any zeros of $\C$. I will now
examine the cosmological evolution near the special points mentioned
above in more detail.

Suppose that $d\C/d\phi=0$ for a finite value of $\phi$ and that
$\C$ is non-singular there. Without loss of generality we can assume
this occurs at $\phi=0$ in which case (to leading order)
\be 
\C = 3 n A\kappa^{-2} + B\phi^n \ ,
\label{Capprox}
\ee
where $A$ and $B$ are constants and $n \geq 2$.
To leading order eq.~\bref{Ceq} implies 
\be
r \sim \left\{ \begin{array}{cc} 
\exp \left(\frac{A}{B(n-2)}\frac{1}{\phi^{n-2}}\right) & n>2 \\ & \\
\phi^{-(A/B)} & n=2 \end{array} \right.
\ee
Thus if $A/B >0$, $r \to \infty$ as $\phi \to 0$, while if $A/B <0$, 
$r \to 0$ instead.

Substituting these approximations into the equation for $U$ \bref{Ueq}, gives
\be
U = -\frac{\kappa^4}{36}\alpha \C^2 
+ \frac{\C}{r^4} + \frac{1}{r^4} \left\{ \begin{array}{cc} 
O\left((\ln r)^{-\frac{n}{n-2}}  \right) & n>2 \\
O\left(r^{-2B/A} \right) & n=2 \end{array} \right.
\ee
where $\alpha$ is an arbitrary integration constant.

Combining these expressions with the symmetric bulk Friedmann equation
\bref{Fried3} gives the late time (large $a$ and $r$) evolution
\be
\left(\frac{\dot a}{a}\right)^2 = \frac{\kappa^4}{36}\left(
\lambda^2-\alpha + 2\lambda \rho + \rho^2\right)\C^2 + \frac{\C}{a^4}
+ o\left(\frac{1}{a^4}\right)
\label{Fried4}
\ee
If we want a non-vanishing effective gravitational constant 
and a vanishing effective cosmological constant,
we require $\C(a \to \infty) \neq 0$ and $\alpha=\lambda^2$. The
above equation then tends to the usual brane cosmology Friedmann
equation, including a so-called dark radiation term. Disappointingly
there are no quintessence-like ($1/r^c$ with $c<2$) or dark matter-like
($1/r^3$) contributions, so this type of solution cannot explain the
observed acceleration of the universe. 

If instead $A=0$ in eq.~\bref{Capprox}, then we are near a zero of
$\C$. In this case 
\be
r \sim \exp\left(-\frac{\kappa^2 \phi^2}{6 n}\right)
\ee
for any $n > 0$, so $\phi=0$ is a local maximum of $r$.

Eq.~\bref{Ueq} then implies
\be
\frac{U}{\C^2} \sim B^{-1} \left\{ \begin{array}{cc} -\phi^{2-n} & n > 2 \\ 
		\ln |\phi| & n=2 \end{array} \right.
\label{Uzapp}
\ee
Note that in order for the point $\phi=0$ to be reachable at all, 
$B$ must be negative.

Combining the Friedmann equation~\bref{Fried3} with eq.~\bref{Ceq} gives
the evolution of $\phi$ with respect to time
\be
\dot \phi^2 = 4\left( \frac{36}{\kappa^4}\frac{U}{\C^2} +
(\lambda+\rho)^2\right)\left(\frac{d\C}{d\phi}\right)^2 \ .
\label{dphidt}
\ee
For $n \geq 2$ the dominant contribution to the first factor on the
right-hand side of eq.~\bref{dphidt} comes from $U/\C^2$
\bref{Uzapp}. Using this approximation, integration of the above
expression reveals that $t \to \infty$ as $\phi \to 0$. Thus it takes an
infinite amount of time to reach the zero of $\C$. As it is approached,
$r$ tends to a constant.

If $n=1$ the dominant part of the right-hand side of eq.~\bref{dphidt}
is generally a constant. $\phi$ thus passes through the zero of $\C$ in
finite time. As $\phi$ passes this point, the universe stops expanding
and begins to collapse. This case is not the same as the re-collapse
near the start of this section, since $\dot \phi$
no longer becomes zero, and the subsequent evolution of the universe is
not just the evolution up to $\dot a =0$ in reverse.

As $\C \to 0$ (for any $n$) the
effective gravitation constant ($8\pi G = \C^2 \kappa^4 \lambda/6$) also
tends to zero. Clearly this type of solution is not compatible with the
standard cosmology. The solution~(\ref{Csol0},\ref{Usol0}), which
corresponds to an exponential potential for $k=0$, also has $\C \to 0$,
although in this case that occurs as $r \to \infty$.

Using eqs.~(\ref{Ceq}--\ref{Veq}) and the components (\ref{GTT},\ref{Grr})
of the Einstein tensor can be written (for any $k$) as
$G^T_T=-6(U + rU'/4)$ and $G^r_r=-6(U-\C/r^4)$. If we require these
to be non-singular, $U$ and $\C$ must be finite for $r>0$. As $r \to 0$
we need $\C \to \const + O(r^4)$ and $U = \C/r^4 + O(1)$. Note that $\C$
cannot tend to zero at $r=0$ since eq.~\bref{Ceq} implies $\C'/\C \leq 0$, 
and so $\C$ tends to a constant or diverges as $r \to 0$. However the
point $r=0$ is still singular, since this behaviour of $\C$ and $U$ implies
$R_{\mu\nu\lambda\rho} R^{\mu\nu\lambda\rho} \sim \C^2/r^8$. Thus there
is always a curvature singularity at $r=0$ unless $\C \equiv 0$. Also,
because $h \sim 1/r^2$ the singularity is a finite proper distance away
from the brane. The only way to avoid having a naked bulk singularity
which is visible from the brane is to hide it behind an event
horizon. At late time (large $a$, $r$) we want $U$ to be negative, so if
there is to be an event horizon, $\C$ must be chosen so
that $U$ is positive at $r=0$. If $\C$ tends to a constant as $r \to 0$,
it needs to be positive to give a suitable black hole solution.

Thus the only viable cosmological solutions with $\C \neq 0$ have bulk
black holes. Our universe then starts at the initial space-like
singularity~\cite{globalst}, which is either an infinity or a maximum of
$|\C|$. If our universe is to expand forever, it must not pass through
any points for which the Friedmann equation~\bref{Fried2} gives $\dot
a =0$. As the scale factor approaches infinity, the value of $\phi$ on
our brane approaches a minimum of $|\C|$. If there are points with $\dot
a = 0$, the universe will either take an infinite time to reach them, or
when they are reached the universe will start to re-collapse. Such points
exist if $\C$ has a zero, or if the effective cosmological constant term of
the Friedmann equation happens to cancel the other terms.

\section{Conformally Flat Solutions}
\label{s:sg}

The above rearrangement of the field equations do not work if $\C \equiv 0$.
This is only possible if $\phi' =0$, in which case the
solution~\bref{stdBW} applies, or if $k=0$. In the second case the
definition of $\C$~\bref{Cdef} implies that $f^2 h = r^2$, and spacetime
is conformally flat. It is useful to set $U=-W^2$, in which case the
field equations imply
\be
-\frac{\kappa^2}{3} r \phi' = \frac{1}{W} \frac{dW}{d\phi} \ .
\label{SGphi}
\ee
\be
V(\phi) = \frac{9}{2\kappa^4} \left(\frac{dW}{d\phi}\right)^2 
- \frac{6}{\kappa^2}W^2 \ , 
\label{SGV}
\ee
The form of $V$ in this case resembles that of supergravity theories,
with $W$ playing the role of the superpotential. The cosmology of this
type of model with an exponential $W$ has been considered
elsewhere~\cite{SGanne}. 

When $\C \equiv 0$ the Friedmann equation \bref{Fried3} becomes
\be
\left(\frac{\dot a}{a}\right)^2 = \left\{-1 + \frac{\kappa^4}{36}\left(
\lambda^2 + 2\lambda \rho + \rho^2\right)\right\}W^2
\ee
which is similar to eq.~\bref{Fried4}, but without the `dark radiation' term.

The equation determining $\phi(r)$~\bref{SGphi} in this case is also
very similar to the corresponding $\C \neq 0$ equation~\bref{Ceq}. Again
the start ($r=a=0$) of our universe occurs at a maximum or infinity of
$|W|$. The field $\phi$ then rolls down $|W|$ until it reaches either a
minimum, in which case $a \to \infty$, or a point with $\dot a =0$. If
$\dot a \to 0$ the universe will either take an infinite amount of time
to reach $\dot a =0$, or the universe will start to re-collapse after
passing this point. The analysis is virtually identical to that of the
previous section. Again we cannot have quintessence-like terms at late
time if we require a zero effective cosmological constant and a
non-vanishing gravitational coupling.

The implications of singularities are different to the $\C \neq 0$
case. Since the metric is conformally flat it is no longer possible to
hide singularities behind an event horizon. A non-singular
energy-momentum tensor implies that $W$ must be finite for all values of
$r$. This implies that $h \sim r^2$ for small $r$, and so the point
$r=0$ is always an infinite proper distance away from the brane. Thus
there are no black holes or naked singularities in this case.

\section{A Cosmologically Viable Example}
\label{s:eg}

The results of the previous two sections suggest that we would like a
model with a $\C$ or $W$ that has a maximum and a minimum and no zeros
between them. In this section I will describe one such model, whose
analytic form allows the equations (\ref{Ceq}--\ref{Veq}) and
(\ref{SGphi},\ref{SGV}) to be integrated relatively painlessly.
For convenience we will change variables to $\varphi = \kappa \phi/\sqrt{3}$.
If
\be
\C = A_1 r_*^4 \left(n(n+1) - \varphi^2\right) e^{\varphi^2/n} \ ,
\label{egC}
\ee
then eq.~\bref{Ceq} implies
\be
\frac{r^4}{r_*^4} 
= \left(\frac{n^2}{\varphi^2} - 1\right)\frac{1}{\varphi^{2n}} \ .
\label{egr}
\ee
Thus $r=0$ at $\varphi=n$, which is a maximum of $\C$, and $r \to \infty$ 
as $\phi \to 0$, where $\C$ has a minimum. Eq.~\bref{Ueq} implies
\bea
&& U = A_1 \left(n(1+n)-\varphi^2\right)^2 e^{\varphi^2/n} 
\left\{ n^n e^{\varphi^2/n} 
\Gamma\left(n,\frac{\varphi^2}{n}\right) 
+\frac{n \varphi^{2n}}{n^2-\varphi^2} \right\}
\nonumber \\ && \hspace{.5in} {}
-A_2 n^n \Gamma(n) \left(n(1+n)-\varphi^2\right)^2 e^{2\varphi^2/n} \ .
\label{egU}
\eea
$A_1$, $A_2$ and $r_*$ and are arbitrary constants.
Here $\Gamma$ is the incomplete gamma function. For positive integer
values of $n$, $e^z \Gamma(n,z) = \Gamma(n) \sum_{k=0}^{n-1} z^k/k!$, and
so the first term of \bref{egU} reduces to $e^{\varphi^2/n}$ times a
polynomial.

To get a vanishing late time effective cosmological constant in the
Friedmann equation~\bref{Fried2} we need $A_2 > A_1$ and $\lambda^2 =
(A_2-A_1) n^n \Gamma(n)/(A_1 r_*^4)^2$. In order to have a black hole
solution as opposed to a naked singularity, we also want $U$ to be positive
as $r \to 0$, and so we need $A_1 > 0$. The quantity $\C$ is then positive for
all values of $r$.

The form of the potential can be obtained from eq.~\bref{Veq}. If $n$ is
an integer it takes the form of a $2n$-th order polynomial multiplied by
$e^{\varphi^2/n}$, plus a sixth order polynomial times
$e^{2\varphi^2/n}$. For example if $n=1$
\be
\kappa^2 V = 6 A_1 (\varphi^2-4)e^{\varphi^2}
-6 A_2 (\varphi^3 + \varphi^2 - \varphi -2)(\varphi^3 - \varphi^2 - \varphi +2)
e^{2\varphi^2}
\label{Veg}
\ee
This would be bounded from below if $A_2 \leq 0$, but as is mentioned
above, we need $A_2$ positive to avoid singularity problems.

The corresponding $\C=0$ solution has $W$ proportional to eq.~\bref{egC}, and
$V$ is given by eq.~\bref{Veg} with $A_1=0$ and $A_2<0$. The
evolution of $r$ is still given by eq.~\bref{egr}. There is no black
hole in this case, but no naked singularity either.

\section{Summary}
\label{s:sum}

To fully understand the behaviour of brane cosmologies with bulk scalar
fields it is not enough to just consider the fields on the brane. A full
bulk solution is required. This is not generally possible, although
solutions in the simpler case of a static bulk with a brane moving
through it can be found. For a $Z_2$ symmetric brane world with a static
bulk, consistency of the solution imposes restrictions on the type of
theory that can exist on the brane. Only models with a very special type
of coupling between the matter fields and the bulk scalar are compatible
with the required boundary condtions. We find that in general the standard
energy momentum condition cannot apply for these solutions. However,
reasonable models which do give the required non-standard conservation
law do exist.

We find from the bulk field equations that the variation of the scalar
field ($\phi$) is directly related to a projection of the bulk Weyl
tensor ($\C/r^4$). The Friedmann equation and the bulk
potential can also be determined from $\C$. The cosmology of the model
is thus determined by the form of the function $\C$ (and the type of
matter present on the brane).

We find that, in general, the cosmology on the brane occurs between two
points with $d\C/d\phi =0$ or $|\C|=\infty$. The universe starts ($a=0$)
at a maximum or infinity of $|\C|$. There are then two possibilities for the
subsequent evolution. Either it can continue to expand as $\phi$ rolls
towards a minimum of $|\C|$ ($a \to \infty$), or $\phi$ can
pass through a zero of $\C$, at which point the universe reaches its
maximum size. After this it contracts until another maximum or infinity
of $|\C|$ is reached, where $a=0$ again. It is also possible for the
evolution to end at a point with $\C=0$, in which case the scale factor
$a$ will asymptotically approach a finite value. 

All the above solutions have singularities at $r=0$, although the
spacetime can have a black hole-like structure, in which case the singularity
is concealed by an event horizon.
If the projected bulk Weyl tensor is zero, the system requires a slightly
different analysis, although the possible cosmologies are very
similar. The main difference is that black hole solutions are no longer
possible. However the point $r=0$ is no longer automatically singular.

The unusual effective energy-momentum conservation equation of these
models, which is required for a consistent solution, implies that the
effective gravitation coupling is proportional to the function
$\C$. Thus solutions which have $\C$ approaching zero are unlikely to be
compatible with the standard cosmology. Requiring a non-vanishing
gravitational coupling has other implications. If we also require an
effective cosmological constant which vanishes at late time, then the
form of the solution implies that there will be no quintessence-like
terms in the Friedmann equation arising from bulk effects.

The only models which agree with the standard cosmology at late times
have $\phi$ rolling down to a minimum of $|\C|$. The late time behaviour
of all such solutions closely resembles that of the well known brane world
cosmologies which do not have scalar fields. Thus if we keep the bulk static
and $Z_2$ symmetric, then the introduction of
scalar fields will not help resolve late time cosmological problems,
such as the observed but unexplained acceleration of the universe.

\section*{Acknowledgements}
I wish to thank Lev Kofman, David Wands, Anne Davis, Philippe Brax,
David Langlois, and Pierre Bin\'etruy for useful comments. I also wish
to thank Ruth Gregory and Christos Charmousis for helpful discussions.

\end{document}